\documentclass{PoS}

\usepackage{lineno}

\title{Constraining the Diffusion Coefficient with HAWC TeV Gamma-Ray Observations of Two Nearby Pulsar Wind Nebulae}

\ShortTitle{Constraining the Diffusion Coefficient with HAWC}

\author{\speaker{Hao Zhou}\\
        Los Alamos National Laboratory\\
        E-mail: \email{hao@lanl.gov}}

\author{Rubén López-Coto\\
        Max-Planck Institute for Nuclear Physics \\
        E-mail: \email{rlopez@mpi-hd.mpg.de}}

\author{Francisco Salesa Greus\\
        The Henryk Niewiadomski Institute of Nuclear Physics, Polish Academy of Sciences\\
        E-mail: \email{fsalesa@ifj.edu.pl}}
        
\author{For the HAWC Collaboration\thanks{For a complete author list, see http://www.hawc-observatory.org/collaboration/icrc2017.php}}

\abstract{

Nearby electron/positron accelerators, mostly Pulsar Wind Nebulae (PWNe), have been proposed as potential origins of the positron excess above 10 GeV. The HAWC Observatory reveals two very extended sources spatially coincident with two nearby middle-aged pulsars: Geminga and PSR B0656+14, suggesting ultra-relativistic electrons/positrons accelerated in our backyard. Morphological studies on these two PWNe provide a constraint on the diffusion coefficient at HAWC energies. In this poster, we will present the model development and morphological studies on these PWNe, and the derived diffusion coefficient that best fits the data.

}

\FullConference{35th International Cosmic Ray Conference --- ICRC2017\\
		10--20 July, 2017\\
		Bexco, Busan, Korea}

\begin{document}

\section{Introduction}

Launched into orbit in 2006, the PAMELA detector discovered an excess in the positron fraction
at energies above 10 GeV as compared to theoretical models of positron production \cite{pamela}. This
anomalous observation has been confirmed with high precision by Fermi Large Area Telescope
(Fermi-LAT) \cite{fermi_e+} and Alpha Magnetic Spectrometer (AMS) \cite{ams}. It was proposed that this
overabundance of positrons could be a consequence of the annihilation or decay of dark matter,
but an alternative explanation is that the positron excess is due to nearby electron/positron
accelerators. Pulsar wind nebulae (PWNe), known as efficient electron/positron accelerators,
were postulated as sources of the positron excess \cite{yuksel}\cite{hooper}. At 250 pc and 288 pc, Geminga (PSR
J0633+1746) and PSR B0656+14 are two of the nearest pulsars to earth, and this proximity
combined with their relatively advanced age make them important candidates contributing to the
locally measured electron and positron flux.

Ultra-relativistic electrons and positrons cool down via inverse Compton scattering (ICS) and
synchrotron. TeV gamma ray can be produced through IC scattering off lower energy photons, e.g. cosmic
microwave background .  The extended TeV PWN with a size of $2.8^\circ \pm 0.8^\circ$ around Geminga was reported by the Milagro experiment \cite{milagro}. Weak evidence ($2.2\sigma$) of this large nebula emission was reported
by the Tibet air shower array \cite{tibet}, but IACT observations using standard analysis techniques
have only provided upper limits. In Fermi-LAT data, the Geminga pulsar is
one of the brightest sources in the GeV sky but there is no unambiguous evidence of the
existence of a surrounding nebula at GeV energies. 

In this contribution, we will demonstrate the method of morphological analysis on the extended TeV emission around Geminga and PSR B0656+14 in order to constraint the particle diffusion and to estimate the contribution from these sources to the local flux of electrons and positrons measured near earth.

\section{Diffusion Coefficient}

\begin{figure}[htpb]
\begin{center}
\includegraphics[width=0.7\linewidth]{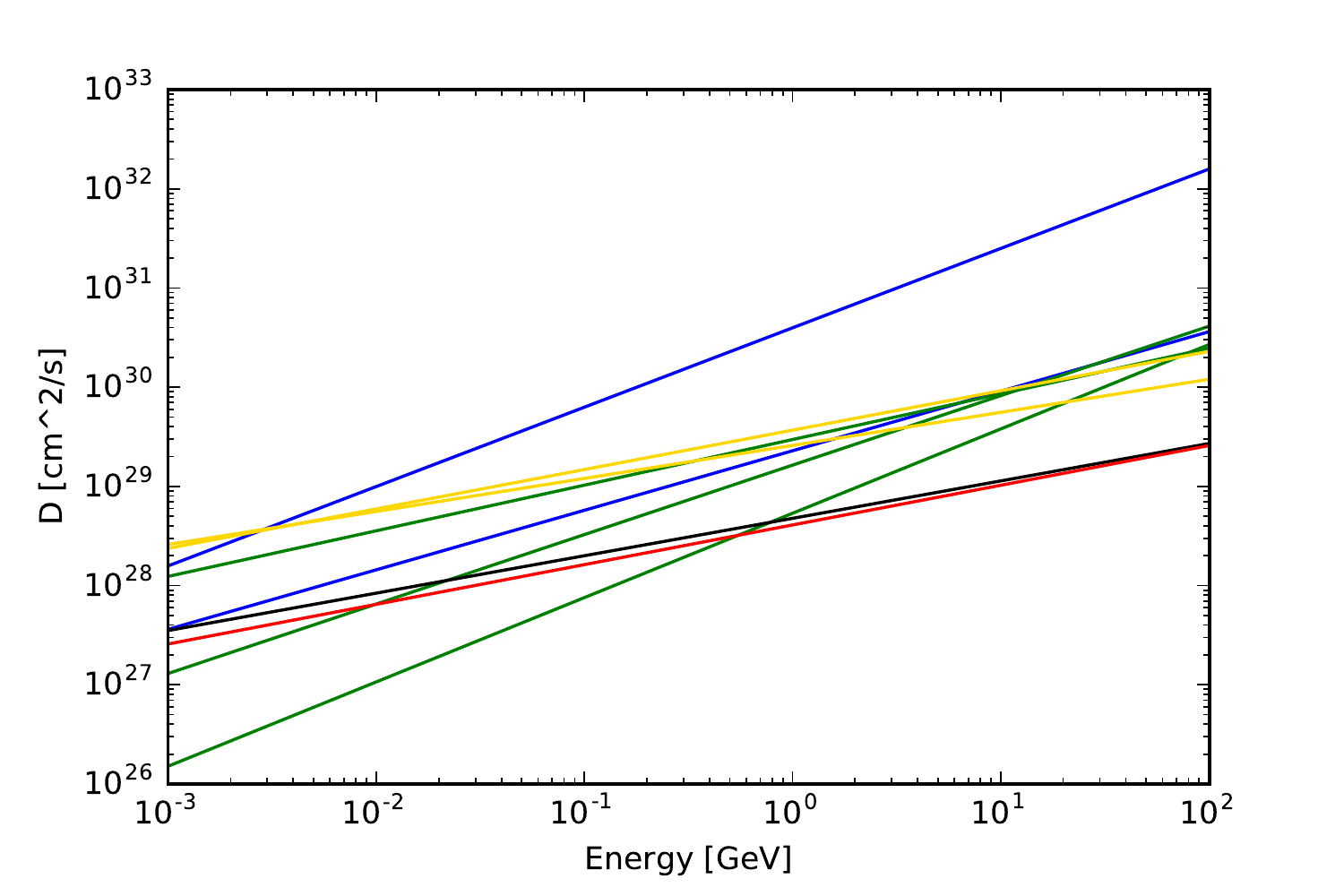}
\caption{Diffusion coefficients from different measurements: blue \cite{strong}, green \cite{delahaye}, black \cite{yin}, red \cite{yuksel}, and yellow \cite{adriani}.}
\label{fig:dcbc}
\end{center}
\end{figure}

Both Geminga and PSR B0656+14 are middle-aged pulsars, with relatively low magnetic field, and the modulation to the surrounding interstellar medium (ISM) due to the particle acceleration by these sources is low. Therefore we consider the scenario that the accelerated particles diffuse isotropically into the ISM. In order to constrain the electron and positron flux that reach the Earth from these two pulsars, we need to know how fast these particles diffuse in the ISM, referred as diffusion coefficient. It is a property of a medium and depends on the energy of the particles,

\begin{equation}
D(E_e) = D_0(E_e/10\,\textrm{GeV})^\delta
\label{equ:dc}
\end{equation}
where the typical value of the diffusion index $\delta$ is 1/3 and $D_0$ is the diffusion coefficient at 10\,GeV. The diffuse coefficient has been measured from the Boron-Carbon ratio in hadronic cosmic rays. Figure \ref{fig:dcbc} summarizes the diffusion coefficients as a function of energy from different measurements.

However, the diffusion coefficient measured from the Boron-Carbon ratio is the average value encountered over very long lifetime of hadronic cosmic rays which are expected to have spent much of that time in the Galactic halo. The local diffusion coefficient could be different. The measurement on the source size around these two nearby pulsars will provide a constraint on the local diffusion coefficient in the ISM.

\section{HAWC Observations}

\begin{figure}[htpb]
\begin{center}
\includegraphics[width=0.5\linewidth]{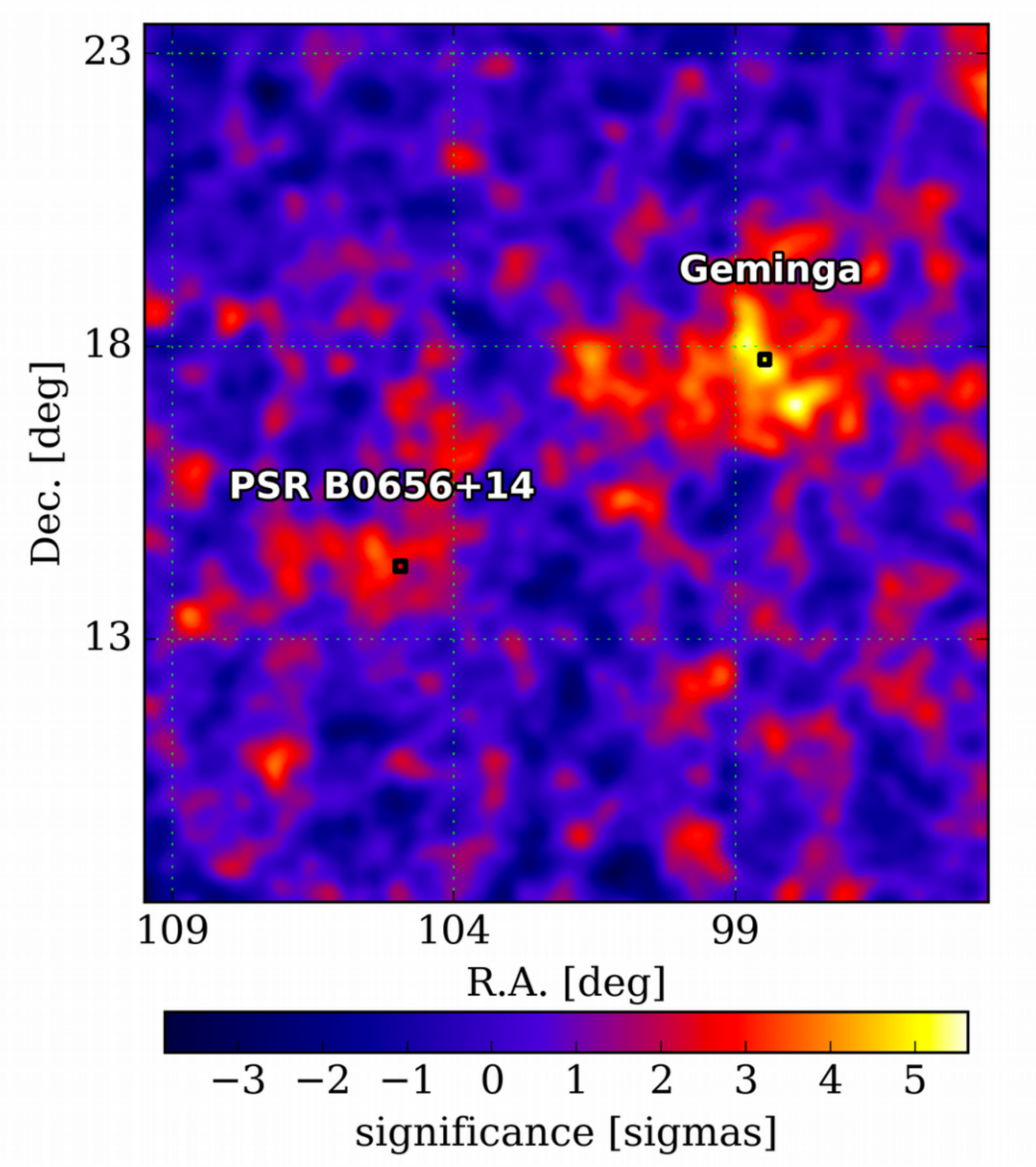}
\caption{The significance map convolved with point spread function in the region around Geminga and PSR B0656+14 with 17 months of HAWC data.}
\label{fig:map}
\end{center}
\end{figure}

The High Altitude Water Cherenkov (HAWC) Observatory, located in central Mexico at 4100 m
above sea level, is sensitive to gamma rays between 100 GeV and 100 TeV. Thanks to its large
field of view of 2 steradians, HAWC has a good sensitivity to extended sources such as pulsar
wind nebulae. With 17 month of data, two very extended sources are detected spatially coincident with Geminga and PSR B0656+14 \cite{2hwc}. Figure \ref{fig:map} shows the significance map in the region around these two pulsars convolved with point spread function.

From Figure \ref{fig:map}, there is clearly extended emission beyond the point spread function of HAWC around both pulsars, showing evidence of electrons and positrons diffuse away from the source into the ISM. We can use the size of the extended sources to constrain the diffusion coefficient. On the studies of extended sources, the most commonly used morphological models are disk and Gaussian models. However, with these models, there is no direct connection from the model parameters to the physical parameters. In this work, we develop a particle diffusion model and apply it to the HAWC data.

\section{Diffusion Model}

The TeV gamma-ray morphology is determined assuming a model where electrons and positrons diffuse isotropically away from the source into the ISM. They produce TeV gamma rays through ICS off low energy photons in the ISM, i.e. cosmic microwave background (CMB), infrared, and optical photons. We hereafter obtain an approximated formula for the gamma-ray emission of the diffusion electron positron cloud which we will use to constrain the size of both HAWC sources.

In case of continuous injection of electrons (and positrons) from a point source at a constant rate $Q_0 E_e^{-\Gamma}$, the radial distribution of the electrons with energy $E_e$ at an instant t and distance r from the source is given by equation 21 of \cite{diffusion},

\begin{equation}
f(t,r,E_e) = \frac{Q_0 E_e^{-\Gamma}}{4\pi D(E_e)r} erfc(\frac{r}{r_d})
\label{equ:diffusion}
\end{equation}
where $D(E_e)$ is the diffusion coefficient as a function of electron energy $E_e$ and $r_d$ is the diffusion radius, up to which the electron efficiently diffuse to. They are defined as,

\begin{equation}
D(E_e) = D_0(E_e/10\,GeV)^\delta
r_d = 2\sqrt{D(E_e)t_E}
\label{equ:dr}
\end{equation}
The typical diffusion index $\delta$ is 1/3 and is fixed in this analysis. $t_E$ is the smaller of two timescales: the injection time t (in this case the age of the pulsar) and the electron cooling time $t_{cool}$, which is a function of election energy and target photon energy,

\begin{equation}
t_{cool} = \frac{m_e c^2}{4/3 c \sigma_T \gamma} \cdot \frac{1}{\mu_B + \mu_{ph}/(1+4\gamma \epsilon_0)^{3/2}}
\label{equ:tcool}
\end{equation}
where $\sigma_T$ is the Thomson cross section, $\gamma$ is the Lorentz factor of electrons, $\mu_B = B^2/8\pi$ is the energy density in the magnetic field, and $\mu_{ph}$ is the energy density of target photon field with average energy $\epsilon_0$ per photon. Equation \ref{equ:tcool} takes into the account the energy loss of electrons due to both synchrotron in magnetic field and ICS off low energy photons. For the elections that produce TeV gamma rays through ICS, where the Klein-Nishina effects are important \cite{kn}, the ICS off infrared and optical photons in the ISM is highly suppressed, leaving only CMB photons as important targets. At these energies, the cooling time is much shorter than the age of these two pulsars. Therefore $t_E$ in equation \ref{equ:dr} can be replaced by $t_{cool}$.

\begin{figure}[htpb]
\begin{center}
\includegraphics[width=0.7\linewidth]{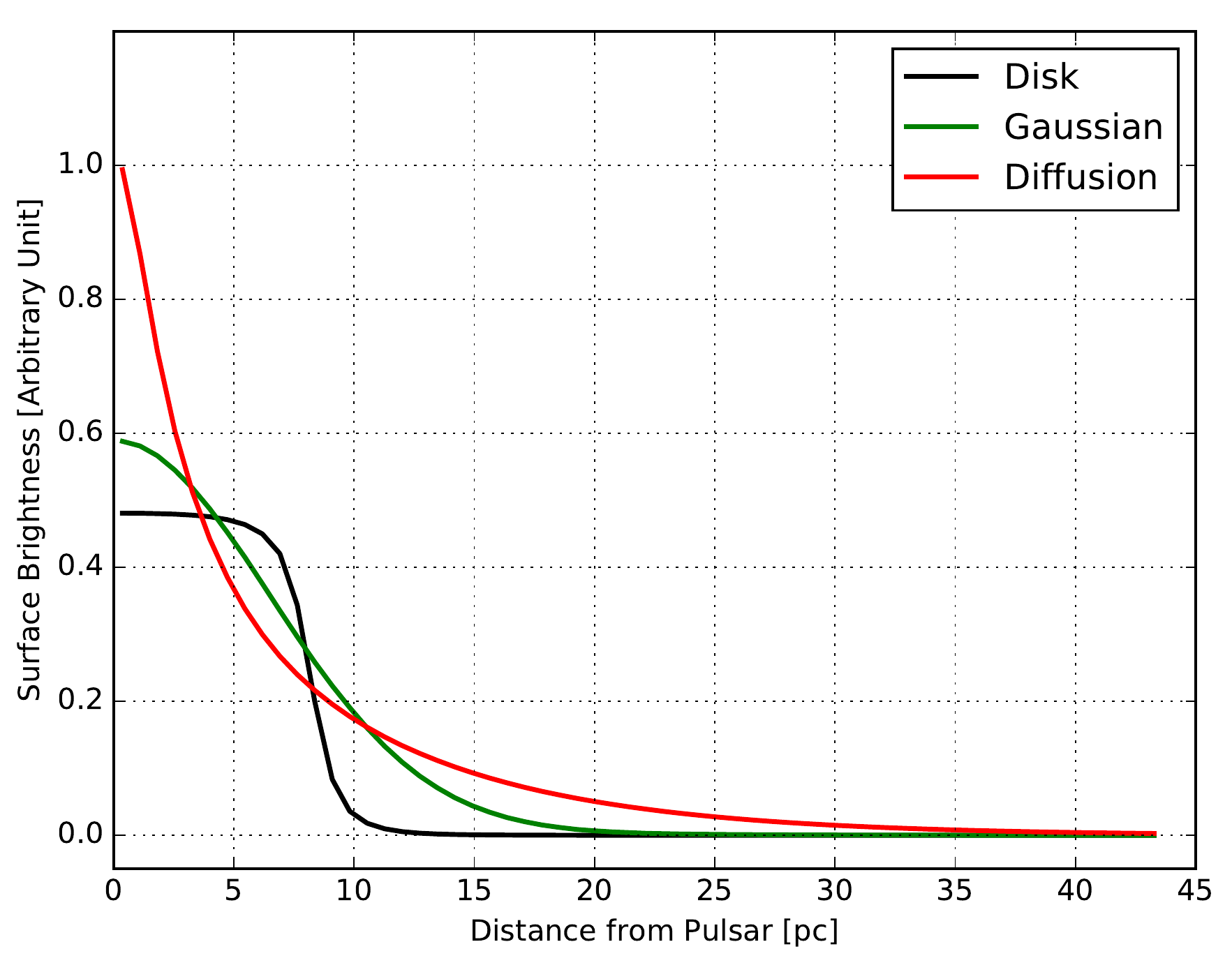}
\caption{Radial profiles of three morphological models: disk, Gaussian, and diffusion model.}
\label{fig:3models}
\end{center}
\end{figure}

Integrating the energy distribution of electrons and positrons along the observer's line of sight, and considering the gamma rays produced through ICS, we obtain the morphology of gamma rays as a function of the distance $d$. An analytical approximation is found on this numerical integral,

\begin{equation}
f_d = \frac{1.2154}{\pi^{3/2} r_d(d+0.06r_d)}exp(-\frac{d^2}{r_d^2})
\label{equ:f2d}
\end{equation}
With a source distance of $d_{src}$, equation \ref{equ:f2d} becomes,

\begin{equation}
f_\theta = \frac{1.2154}{\pi^{3/2} \theta_d(\theta+0.06\theta_d)}exp(-\frac{\theta^2}{\theta_d^2})
\label{equ:f2d}
\end{equation}
where $\theta$ is the angular distance from the pulsar and $\theta_d = r_d/d_{src} \cdot 180^\circ/\pi$ is the diffusion angle. Combined with equation \ref{equ:f2d}, \ref{equ:dr}, and \ref{equ:tcool}, the diffusion angle is a function of electron energy $E_e$ and the target field,

\begin{equation}
\theta_d = \theta_0 (\frac{E_e}{E_{e0}})^\frac{\delta-1}{2}\sqrt{\frac{B^2/8\pi+\mu_{ph}/(1+4\epsilon_0E_{e0}/m_ec^2)^{3/2}}{B^2/8\pi+\mu_{ph}/(1+4\epsilon_0E_{e}/m_ec^2)^{3/2}}}
\label{equ:da}
\end{equation}
where $\theta_0$ is the diffusion angle at the pivot energy of $E_{e0}$. The relation between the mean electron and gamma ray energy is given by \cite{eg},

\begin{equation}
<E_e> \approx 17 <E_\gamma>^{0.54+0.046log_{10}(<E_\gamma>/TeV)}
\label{equ:eg}
\end{equation}
The pivot energy $E_{e0}$ is chosen to be 100 TeV in this analysis, which corresponds to $\sim 20$ TeV gamma rays. Figure \ref{fig:3models} compares the radial profile as a function of the distance from the pulsar from three morphological models.

We then fit the gamma-ray emission around Geminga and PSR B0656+14 with the diffusion model defined in equation \ref{equ:f2d} and \ref{equ:da} using the Multi-Mission Maximum Likelihood framework \cite{3ml}, and calculate the diffusion coefficient of 100 TeV electrons based on the obtained diffusion angle from the likelihood fit.

\section{Results and Discussion}

HAWC observations reveal two very extended TeV gamma-ray sources spatially coincident with Geminga and PSR 0656+14, suggesting ultra-relativistic electrons and positrons accelerated in our neighborhood. The results of the analysis on these two sources and the obtained diffusion coefficient will be presented at ICRC2017. The implication on the positron contribution of these sources to the local flux can be found in other two proceedings for this conference \cite{paco} \cite{ruben}.

\section*{Acknowledgments}
We acknowledge the support from: the US National Science Foundation (NSF); the
US Department of Energy Office of High-Energy Physics; the Laboratory Directed
Research and Development (LDRD) program of Los Alamos National Laboratory;
Consejo Nacional de Ciencia y Tecnolog\'{\i}a (CONACyT), M{\'e}xico (grants
271051, 232656, 260378, 179588, 239762, 254964, 271737, 258865, 243290,
132197), Laboratorio Nacional HAWC de rayos gamma; L'OREAL Fellowship for
Women in Science 2014; Red HAWC, M{\'e}xico; DGAPA-UNAM (grants IG100317,
IN111315, IN111716-3, IA102715, 109916, IA102917); VIEP-BUAP; PIFI 2012, 2013,
PROFOCIE 2014, 2015;the University of Wisconsin Alumni Research Foundation;
the Institute of Geophysics, Planetary Physics, and Signatures at Los Alamos
National Laboratory; Polish Science Centre grant DEC-2014/13/B/ST9/945;
Coordinaci{\'o}n de la Investigaci{\'o}n Cient\'{\i}fica de la Universidad
Michoacana. Thanks to Luciano D\'{\i}az and Eduardo Murrieta for technical support.

\end{document}